\documentclass{article}[]

\usepackage{graphics,epsfig}

\begin{document}

\hfill {\footnotesize WM-02-109;  JLAB-THY-02-54}  \vskip 1.5cm

\centerline{\bf INTERPLAY OF HARD AND SOFT PROCESSES AT JLAB ENERGIES\footnote{Written version of invited talk by CEC at the Second International Symposium on the Gerasimov-Drell-Hearn sum rule 
and the spin structure of the nucleon (GDH2002), Genoa, Italy, 3--6 July 2002, and of talk by CEC under the title ``Hard Pion Electroproduction at Medium Energies'' at the XVI International Conference on Particles and Nuclei (PANIC02), Osaka, Japan, 30 September--4 October 2002 [WM-02-111; JLAB-THY-02-62].
}}

\bigskip

\centerline{ANDREI AFANASEV$^*$ and 
{CARL~E. CARLSON}$^{*,\dagger}$}

\medskip

\begin{centering}{\it \small
$^*$Thomas Jefferson National Accelerator Facility,\\
12000 Jefferson Avenue, Newport News, VA 23606, USA\\
and\\
$^\dagger$Nuclear and Particle Theory Group, Department of Physics,\\
College of William and Mary, Williamsburg, VA 23187-8795, USA\\
}
\end{centering}

\medskip




\begin{quote}
{\small
Even at moderate energy machines, there is a regime where hard pion
electroproduction proceeds by a perturbatively calculable process. 
The process, we claim, is not the leading twist fragmentation
one but rather a higher twist process that produces kinematically
isolated  pions. Semiexclusive data may teach us more about parton distribution functions of the target and the pion
distribution amplitude.
In addition, there  is a connection to generalized parton
distribution calculations of exclusive processes in that the perturbative kernel is the same.
}
\end{quote}



\vskip 0.5em

\noindent {\bf 1. Introduction}

\vskip 1em


We are going to discuss semiexclusive photoproduction of hard pions~\cite{photopi,bdhp}, 
\begin{equation}
\gamma + p \rightarrow \pi + X
\end{equation}

\noindent and the semiexclusive deep inelastic scattering version of the same~\cite{inprep},
\begin{equation}
e + p \rightarrow e + \pi + X  \ ,
\end{equation}

\noindent which we can also write as $p(e,e' \pi)X$.  We are interested  in pions with large transverse momentum (that is what ``hard'' means), and particularly in pions that are kinematically isolated, rather than pions that are part of a jet.  And further, we shall hope to isolate processes that can be calculated perturbatively in the context of Quantum Chromodynamics (QCD).

What is our motivation for this interest?
It is severalfold.  

$\bullet$ The calculated result is sensitive to quark distributions in the target and---if we can find a region where the perturbative calculation is valid---we can use hard pion production to measure the high-$x$ quark distribution.  

$\bullet$ The flavor of the outgoing meson favors particular target quarks, and thus we can choose the flavor quark we wish to  measure.

$\bullet$ At moderate energy machines, the dominant process is a higher twist one (i.e., one which disappears like $1/s$ or $1/Q^2$ at high energies) that is none-the-less perturbatively calculable.

$\bullet$ There is a connection to generalized parton distributions, which are the quantities one hopes to measure in a deep exclusive scattering like $\gamma^* + p \rightarrow \pi + N$.  The basic subprocess is the same as the one for the higher twist contributions to the semiexclusive reaction.  Hence worries about, for example, having sufficient energy and $Q^2$ are related, and semiexclusive and deep-exclusive meson production become relevant to each other.

We will introduce the discussion by talking about the perturbatively calculable leading twist process along with the not perturbatively calculable soft processes in the next section, and then talk about the higher twist process, and its regions of dominance, in the section after.  We will end with a brief summary.



\vskip 2.5em

\noindent {\bf 2. Leading twist processes and soft processes}

\vskip 1em

\begin{figure}   [ht]
\centerline{  \epsfysize 1.3in \epsfbox{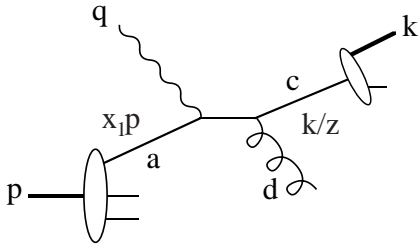}  }
\caption{A fragmentation process, which is leading twist}  \label{frag}
\end{figure}

At high energy and at high (but not superhigh, as it will turn out) pion transverse momentum, hard pion photo- or electroproduction is dominated by the leading twist process or fragmentation process.  A number of subprocesses  contribute, and a typical one is shown in Fig.~\ref{frag}.


The photon, real or virtual, strikes a quark. The quark carries momentum fraction $x_1$ of the target, and the distribution in $x_1$ is described  by a parton distribution function.  The observed pion comes from the fragmentation of an outgoing parton, quark ``c'' in the figure, and the pion carries momentum fraction $z$ of its parent parton.  The distribution in $z$ is described by a fragmentation function.  Since there is a hard  subprocess, the reaction is calculable in perturbation theory.  Comparison of calculation with data taken at high energy machines shows excellent agreement, for pion transverse momenta $k_\perp$ above about 2 GeV.

Below about 2 GeV (the number is not fixed, but depends on the incoming energy and the angle at which the pion emerges), there are important soft processes that are not calculable {\it ab initio}.  This leads to the worry that at moderate initial energies, where the kinematic limit on $k_\perp$ may not be much above 2 GeV, we cannot calculate anything observable.  

\begin{figure}   [ht]
\centerline{  \epsfysize 1.1in \epsfbox{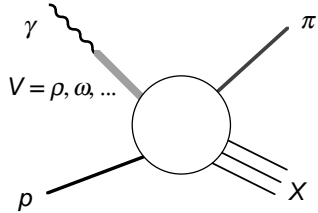}  }
\caption{A soft process}  \label{two}
\end{figure}

To see if this is so, we can estimate phenomenologically the soft processes.  The soft process proceeds mainly by vector meson dominance, (VMD), wherein the entering photon fluctuates into rho or other vector  meson which then interacts with the  target, as illustrated in Fig.~\ref{two}.


The photon amplitude is related to the rho meson amplitude by
\begin{equation}
\left. f(\gamma p \rightarrow \pi^+ X) \right|_{\rho\rm MD}
  = {e\over f_\rho} f(\rho^0 p \rightarrow \pi^+ X)
\end{equation}

\noindent or for the cross section,
\begin{equation}
d\sigma (\gamma p \rightarrow \pi^+ X) 
  = {\alpha \over \alpha_\rho } 
    d\sigma (\rho^0 p \rightarrow \pi^+ X)
                                          \nonumber \\[1ex]
    + {\rm other\ VMD\ contributions} \ .
    \end{equation}
    
\noindent Non-VMD contributions are ignored.  Quantity $e/f_\rho$ is the photon-rho meson coupling, which is gotten from the rate
$\Gamma(\rho \rightarrow e^+ e^-)$, and 
$\alpha_\rho\equiv f_\rho^2/4\pi$.

We need a parameterization of the hadronic process.  There is an excellent one in Szczurek {\it et al.}~\cite{szczurek}, but we have one too, and our results use that one~\cite{soft}.  For information, the hadronic data we actually use is an average for $\pi^+$ and $\pi^-$ initiated  reactions, since data for $\rho^0$ beams is lacking.  Crucial data comes from Bosetti {\it et al.}~\cite{bosetti}, who have pion-in, pion-out semiexclusive data at many angles, but over a limited  energy range.  There is, however, data at many energies for $pp \rightarrow \pi X$ at 90$^\circ$ in the center-of-mass, with a fit given by Beier {\it et al.}~\cite{beier}  Where the data overlap the naive relation
\begin{equation}
E_\pi {d\sigma \over d^3k}
      (\pi^+ p \rightarrow \pi^- X) = {2 \over 3}
E_\pi {d\sigma \over d^3k}
      (pp \rightarrow \pi^- X) 
      \end{equation}

\noindent works well.  Our parameterization~\cite{soft} is a hybrid of a fit to the angular distribution of Bosetti {\it et al.} with the existing 90$^\circ$ energy dependent fit of Beier {\it et al.}, and includes estimates of contributions from other vector mesons, such as the $\phi$, the $\omega$, and vector meson excitations.  

\begin{figure}
\epsfxsize 2.35in \epsfbox{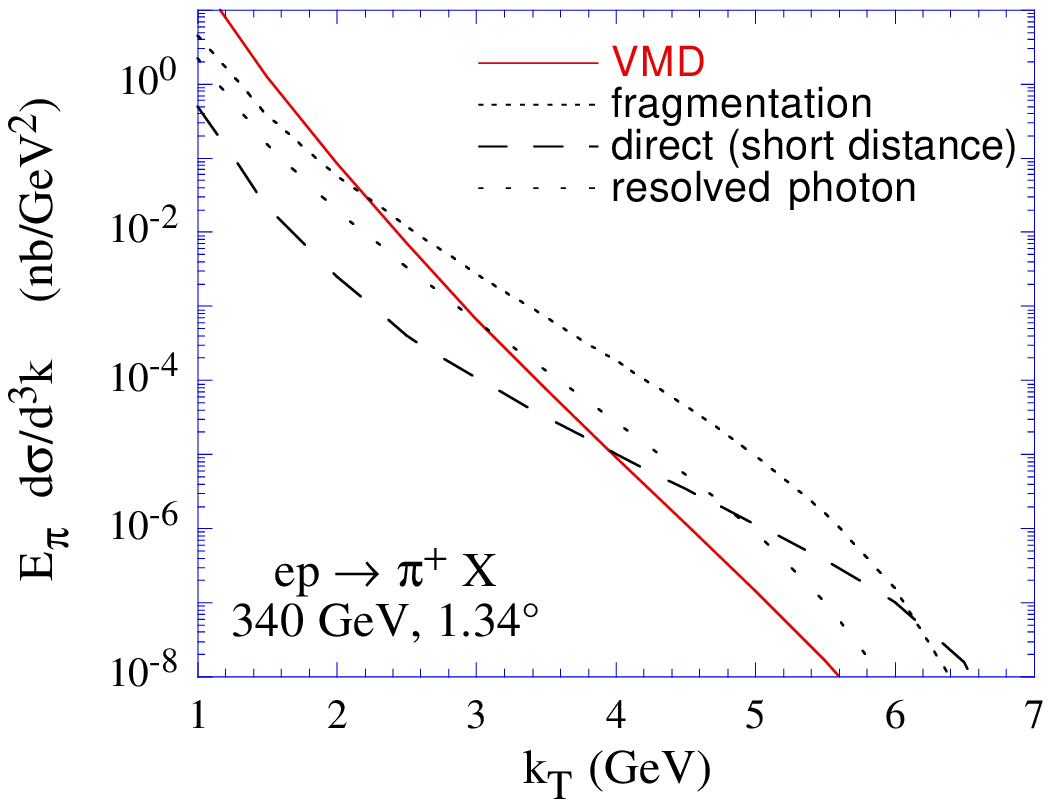}  
\raisebox{-1.7ex}{\epsfxsize 2.35in \epsfbox{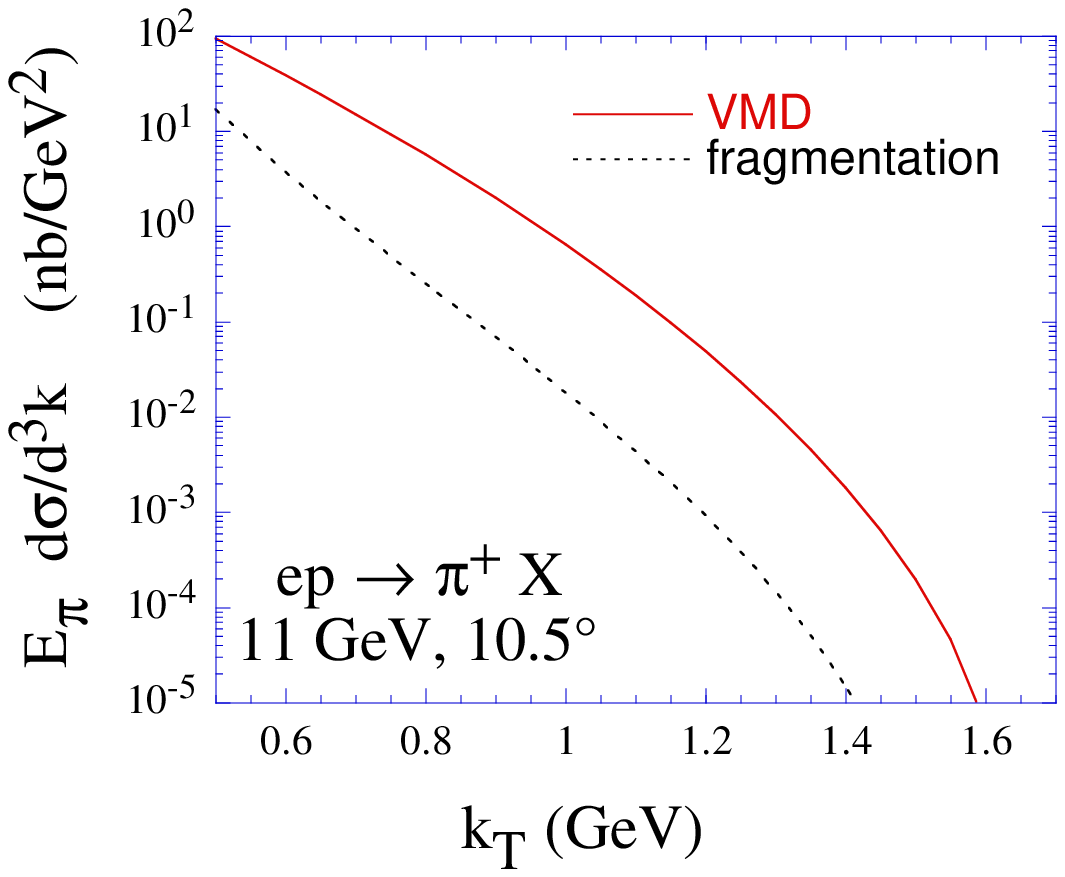}}
\caption{Soft and  fragmentation contributions induced  by real photons.}   \label{real}
\end{figure}

We show two relevant plots in Fig.~\ref{real}.  Both are for real photons, in the sense that for an incoming electron with the outgoing electron not observed, the reactions are due to bremsstrahlung photons with a spread of energies but always close to the mass shell.  


On the left is a higher energy plot (labeled on the figure in the target rest frame, but also corresponding to a 40 GeV electron beam colliding with a 4 GeV proton beam, with pions emerging at 90$^\circ$).  There is a long region where the leading twist process, labeled ``fragmentation'', dominates.  The curve labeled ``direct'' will be discussed soon.  On the other hand, for the 11 GeV plot on the right, the soft processes, labeled ``VMD'', always lie above the leading twist process.  This does not here look hopeful for finding a perturbatively calculable cross section these lower energies~\cite{eden}, but we shall continue in the next section by discussing another perturbative process that is larger than the leading twist one in these regions,  and also discuss how to kinematically suppress the soft contributions.



\vskip 2.5em

\noindent {\bf 3. A higher twist process: direct pion production}

\vskip 1em


The soft process cross section falls off with increasing pion transverse momentum, but the leading twist perturbative cross section falls off also.  There are a number of straightforward reasons for this. The reason we want to call attention to follows from the fact that the pion from leading twist is part of a jet.  The pion has to share momentum with other particles moving in the same direction, and the fragmentation function tells that the probability of finding a pion falls with increasing $z$, typically like $(1-z)^3$ ($z$ is the fraction of the jet momentum carried by the  observed pion).

\begin{figure}   [ht]
\centerline{  \epsfysize 1.3in \epsfbox{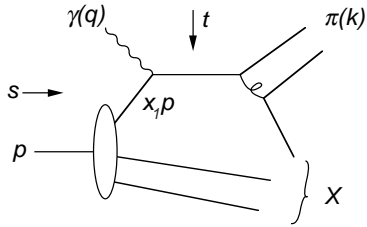}  }
\caption{A process variously called isolated pion production,  direct pion production, or higher twist pion production.}  \label{iso}
\end{figure}

So it is in our interest to examine the process where the pion is produced in kinematic isolation~\cite{photopi,bdhp,baier,berger,milana,wakely,hyer,brand}.  See Fig.~\ref{iso}.  The  color of the outgoing quark must be neutralized right away, and here it is done as the gluon splits into a quark-antiquark pair and a pion is formed at short distance.  


A notable characteristic of this process is that it is proportional to the pion wave function at the origin.  This can be represented by a factor of the pion decay constant $f_\pi$, since the latter is proportional to the  same quantity.  Hence there is a factor like $f_\pi^2/s$ in the direct process cross section, showing that it is higher twist (i.e., power law suppressed at high energy).  However, the direct process does have the feature that it delivers all the momentum in the pion's direction to the one pion.  Hence the process can be significant---possibly the biggest process there is---at the highest $k_\perp$.

\begin{figure}
\epsfxsize 2.35in \epsfbox{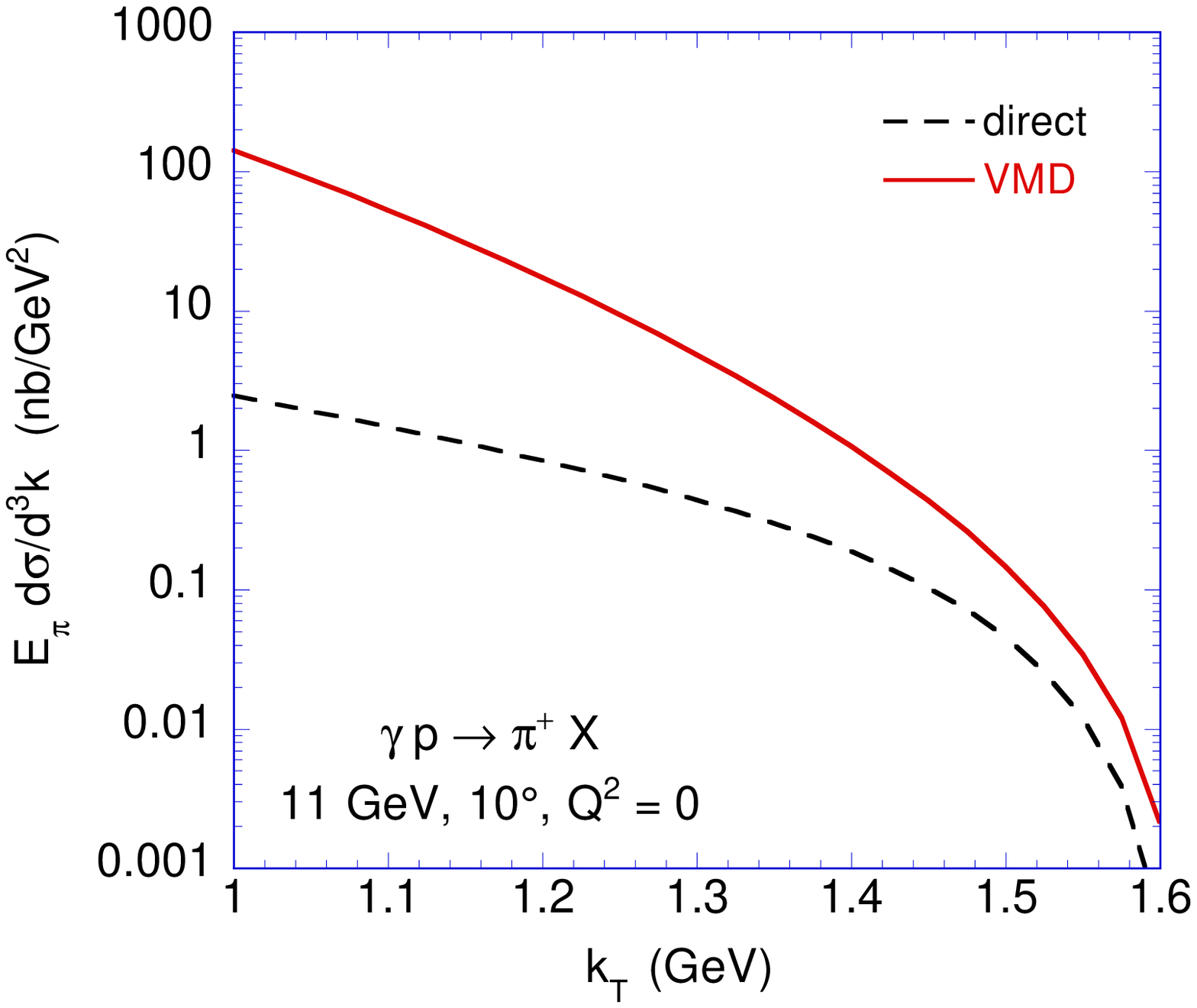}  
\epsfxsize 2.35in \epsfbox{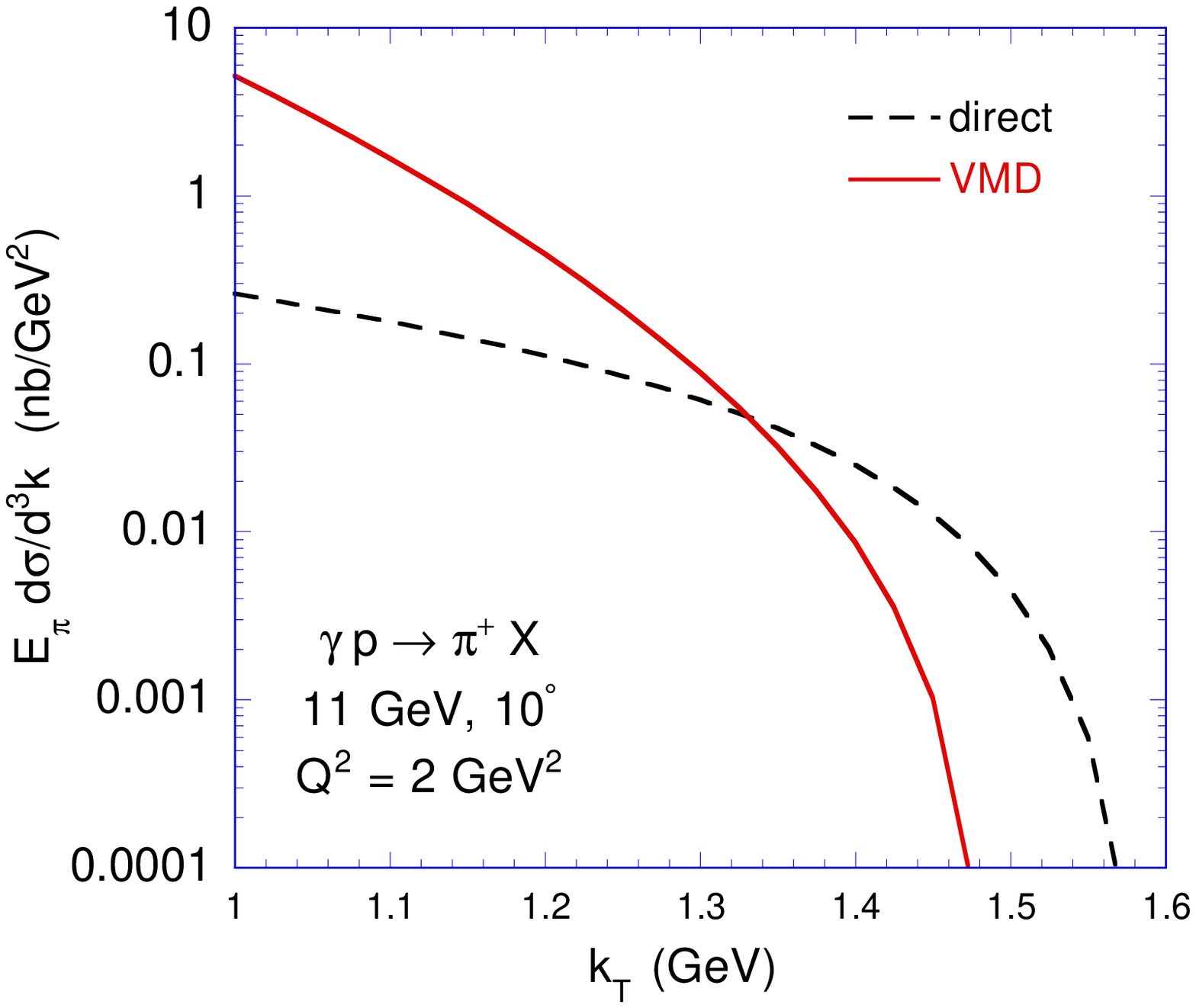}
\caption{On and off shell transversely polarized photon processes.}   \label{offshell}
\end{figure}

We may still want to suppress the soft process further.  There is a way to do so, and that is to put the photon off shell by doing electroproduction with the final electron observed.  Since the soft process is  well approximated  by vector meson dominance, we can see that there is a simple change in the rho (say) meson propagator, so that the amplitude is reduced  by a factor 
$(m_\rho^2/(m_\rho^2 + Q^2))$.  This is numerically a factor of (1/7) in the cross section for $Q^2=1$ GeV$^2$.  The direct (and fragmentation) process is also reduced as the photon goes off-shell, but not by as much.  Hence the importance of the perturbative process increases as the photon goes off-shell.  Some calculated results are shown in Fig.~\ref{offshell}.


These figures show differential cross sections for off-shell transverse photons (for experts, we use the Hand convention to define a flux), with two different values of the photon four-momentum squared.  The leading twist contribution, not shown, is for these kinematics never the biggest process.  The incoming energy is one that is possible with the JLab upgrade.  For on-shell photons, on the left, the soft process is always bigger than the direct  process: not what we want to see.  But, on the right, with $Q^2 = 2$ GeV$^2$, the direct process rises above the soft process for high enough pion transverse momentum, in this case above 1.33 GeV.  

These plots were made with a particular parton distribution function, namely the one of Brodsky, Burkardt, and Schmidt~\cite{bbs}.  There are, of course, others, and they are not the same, particularly at high quark momentum fraction $x$ (or $x_1$ as we called it in Fig.~\ref{iso}) and particularly for the down quark.  One outcome of seeing isolated pion production is to distinguish among the various models for the parton distribution functions.  We should mention that the momentum fraction $x_1$ of the struck quark is directly expressible in terms of measurable quantities.  If one defines the Mandelstam variables from the 3 observed particles, $s=(p+q)^2$, $t=(q-k)^2$, and $u=(p-k)^2$, then one can show that (neglecting masses)
\begin{equation}
x_1 = {-t \over s+u+Q^2}  \ .
\end{equation}

\noindent One can also work out $m_X$, the mass of the  hadronic matter recoiling against the pion. This leads to a ``truth in advertising'' comment.  At the crossover point in the right hand graph of Fig.~\ref{offshell}, the value of $x_1$ is about 1/2 and $m_X$ is about 2 GeV (and decreases as we go farther to the right).  This means that we are in the resonance region, which one can put a positive light by entertaining the idea of checking duality in a semiexclusive process.  At HERMES energies, however, we are safely in the hadronic continuum.  For a numerical example, using 27 GeV entering photons, with $Q^2$ of 2 GeV$^2$ and pions emerging at 7$^\circ$ in the target rest frame, the crossover point is at 1.9 GeV transverse momentum, with $x_1$ about 0.33 and $m_X$ about 4 GeV.

We have discussed transverse photons.  Generally, in electroproduction the photon has a mixture of polarizations which can be separated by the angular dependence, as in
\begin{equation}
{d\sigma \over dE_\gamma \, dQ^2 \, d^3k} \propto
\left(\sigma_T + \epsilon\sigma_L 
+ 2\epsilon \cos{2\phi} \sigma_{TT}
+ \sqrt{2\epsilon(1+\epsilon)} \cos{\phi} \sigma_{LT}
\right)         \ .
\end{equation} 

\begin{figure} [ht]
\epsfxsize 2.35in \epsfbox{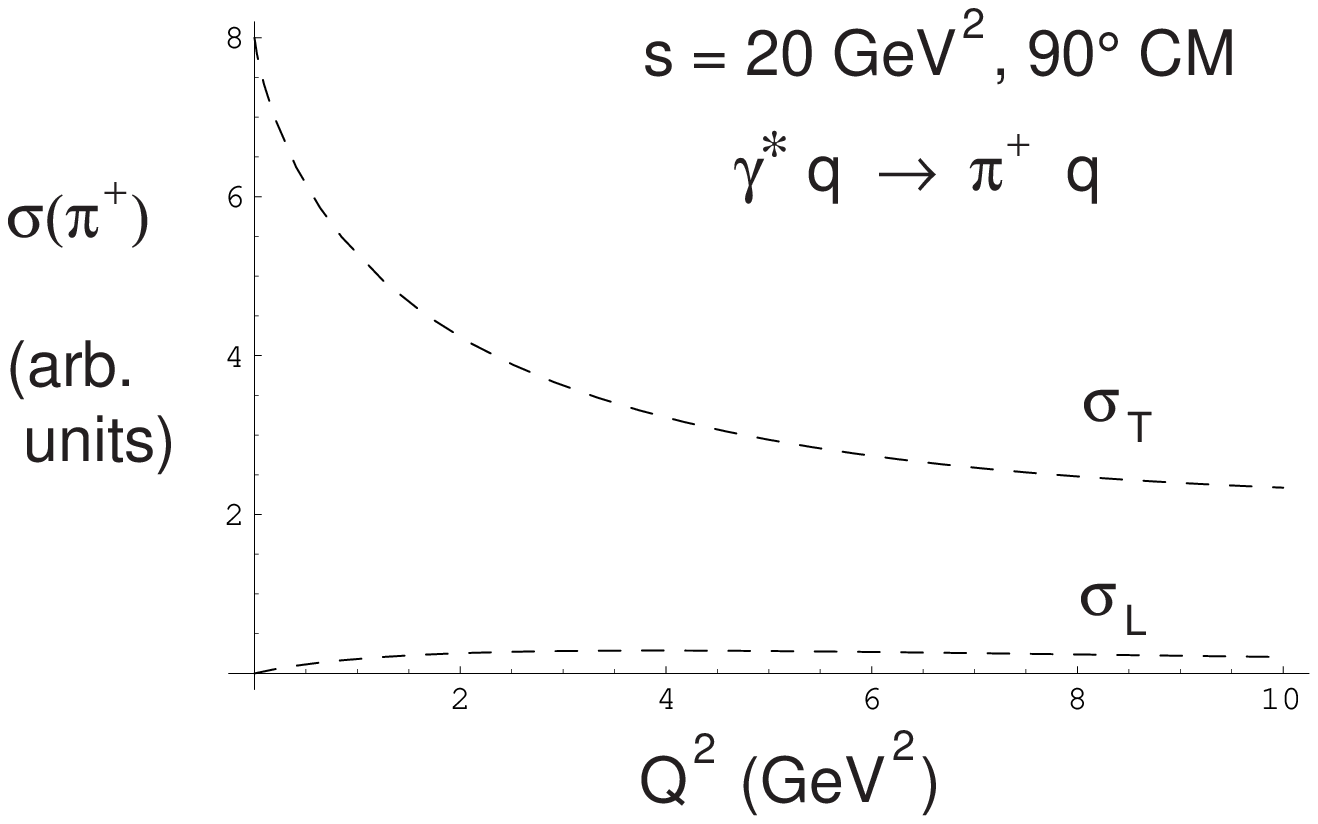}  
\epsfxsize 2.35in \epsfbox{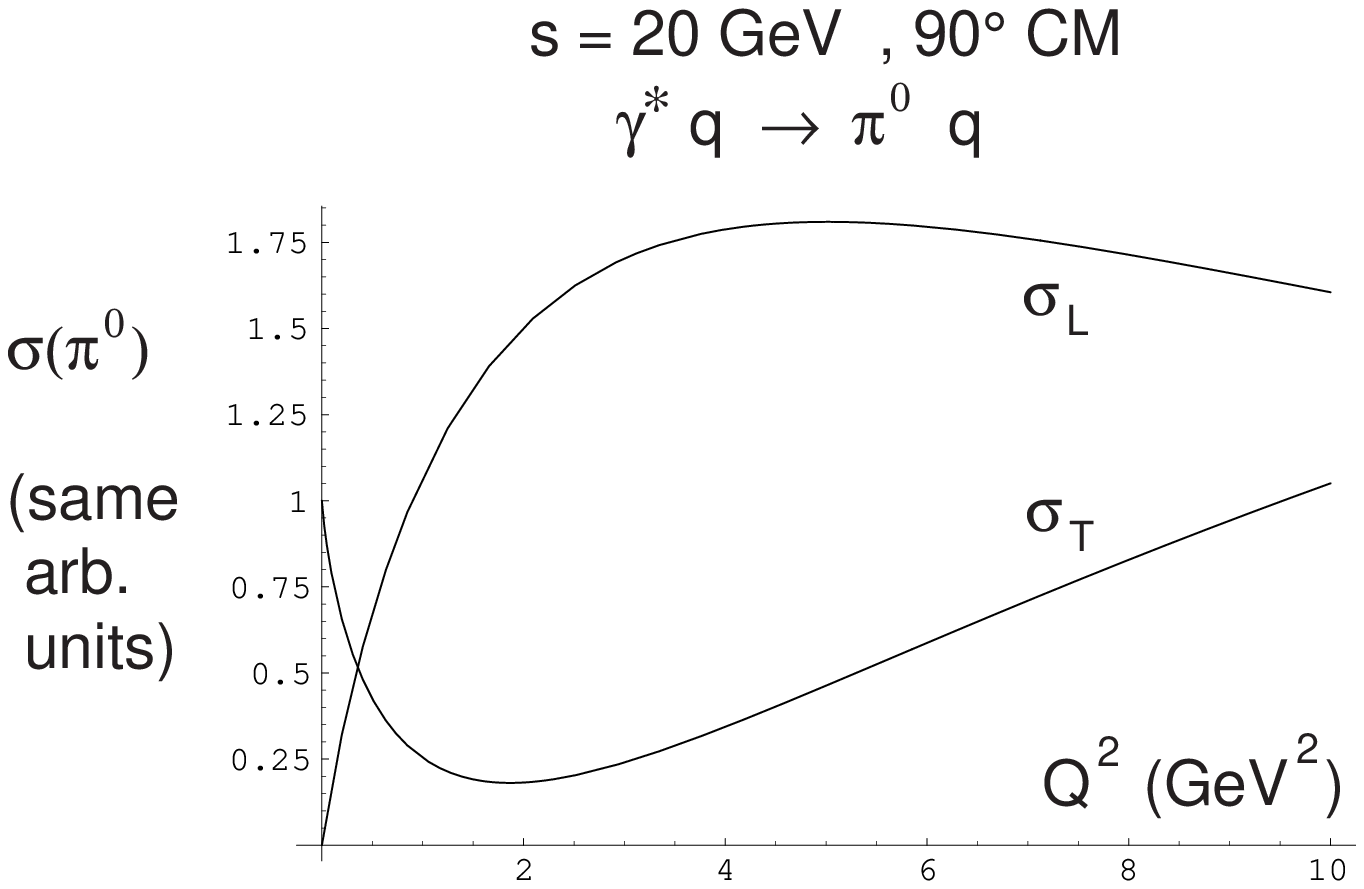}
\caption{$\sigma_T$ and $\sigma_L$ for positive and neutral pions.}   \label{character}
\end{figure}

\noindent Here, $\epsilon$ is the ratio of longitudinal to transverse photon polarization, fixed in standard ways by the photon kinematics, $\phi$ is the azimuthal angle between the electron scattering plane and the photon-pion plane, $\sigma_T$ is the transverse cross section, $\sigma_L$ is the longitudinal cross section, and $\sigma_{TT}$ and $\sigma_{LT}$ are interference terms.  The direct pion production subprocess gives characteristic relations between $\sigma_T$ and $\sigma_L$ which are different for different charge pions and illustrated in Fig.~\ref{character} (there are equivalent figures for $\sigma_{TT}$ and $\sigma_{LT}$).


\vskip 2.5em

\noindent {\bf 4. Summary}

\vskip 1em


We have shown by calculation that even at moderate energy machines, there is a regime where hard pion production proceeds by a perturbatively calculable process.  However, the process, at moderate  energy machines, is not the leading twist fragmentation one but rather a higher twist process that produces kinematically isolated  pions.  To our knowledge, there is no analyzed data that we can compare to, although finding data is possible with present machines.  

Since the process is calculable, and dependent on such things as the
parton distribution functions of the target and the pion distribution
amplitude, one may use the measurement and calculation to learn more
about those quantities, particularly about the parton distributions at
high $x$.   In addition, the perturbative kernel for the semiexclusive
process is the same as in a generalized parton distribution calculation
of exclusive processes, and semiexclusive results may cast light on how
well generalized parton distribution experiments at these energies can
be analyzed.


\section*{\normalsize Acknowledgments}


C.E.C. thanks the NSF for support under Grant PHY-9900657.  Both of us thank the DOE for support under contract DE-AC05-84ER40150 under which the Southeastern Universities Research Association (SURA) operates the Thomas Jefferson National Accelerator Facility.


\begin{thebibliography}{0}

\bibitem{photopi}
A.~Afanasev, C.~E.~Carlson, and C.~Wahlquist,
Phys.\ Lett.\ B {\bf 398}, 393 (1997)
[arXiv:hep-ph/9701215];
Phys.\ Rev.\ D {\bf 58}, 054007 (1998)
[arXiv:hep-ph/9706522].

\bibitem{bdhp}
S.~J.~Brodsky, M.~Diehl, P.~Hoyer, and S.~Peigne,
Phys.\ Lett.\ B {\bf 449}, 306 (1999)
[arXiv:hep-ph/9812277].

\bibitem{inprep} A.~Afanasev and C.~E.~Carlson, in preparation.

\bibitem{szczurek}
A.~Szczurek, V.~Uleshchenko, and J.~Speth,
Phys.\ Rev.\ D {\bf 63}, 114005 (2001)
[arXiv:hep-ph/0009318].

\bibitem{soft}
A.~Afanasev, C.~E.~Carlson, and C.~Wahlquist,
Phys.\ Rev.\ D {\bf 61}, 034014 (2000)
[arXiv:hep-ph/9903493].

\bibitem{bosetti}
P. Bosetti {\it et al.}, Nucl. Phys. {\bf B54}, 141 (1973).

\bibitem{beier}
E.~W.~Beier {\it et al.},
Phys.\ Rev.\ D {\bf 18}, 2235 (1978).

\bibitem{eden}
P.~Eden, P.~Hoyer, and A.~Khodjamirian,
JHEP {\bf 0110}, 040 (2001)
[arXiv:hep-ph/0110297].


\bibitem{baier}
V.~N.~Baier and A.~G.~Grozin,
Phys.\ Lett.\ B {\bf 96}, 181 (1980).

\bibitem{berger}
E.~L.~Berger and S.~J.~Brodsky,
Phys.\ Rev.\ D {\bf 24}, 2428 (1981).

\bibitem{milana}
C.~E.~Carlson and J.~Milana,
Phys.\ Rev.\ D {\bf 44}, 1377 (1991);
Phys.\ Rev.\ Lett.\  {\bf 65} (1990) 1717.


\bibitem{wakely}
C.~E.~Carlson and A.~B.~Wakely,
Phys.\ Rev.\ D {\bf 48}, 2000 (1993).


\bibitem{hyer}
T.~Hyer,
Phys.\ Rev.\ D {\bf 48}, 147 (1993);
Phys.\ Rev.\ D {\bf 50}, 4382 (1994).


\bibitem{brand}
A.~Brandenburg, V.~V.~Khoze, and D.~M\"uller,
Phys.\ Lett.\ B {\bf 347}, 413 (1995)
[arXiv:hep-ph/9410327].

\bibitem{bbs}
S.~J.~Brodsky, M.~Burkardt, and I.~Schmidt,
Nucl.\ Phys.\ B {\bf 441}, 197 (1995)
[arXiv:hep-ph/9401328].




\end{thebibliography}
\end{document}